# Effects of Surface Trapping and Contact Ion Pairing on Ion Transport in Nanopores

Zhongwu Li[1†], Yinghua Qiu[2†], Yan Zhang[1], Min Yue[1], and Yunfei Chen[1]*

1. Jiangsu Key Laboratory for Design and Manufacture of Micro-Nano Biomedical Instruments, School of Mechanical Engineering, Southeast University, Nanjing 211189, China
2. Department of Chemistry, University of Utah, Salt Lake City, Utah 84112, United States

**Abstract**

Ion transport in highly-confined space is important to various applications, such as biosensing and seawater desalination with nanopores. All-atom molecular dynamics simulations are conducted to investigate the transport of $Na^+$ and $Cl^-$ ions through nanopores with the diameter below 6 nm. It is found that the formation of the contact ion pair plays a critical role in reducing the ion mobility inside a nanopore without surface charges. The mobility for both cations and anions decreases with the reduced pore size because it is easier to form the contact ion pairs inside the neutral nanopore with smaller diameter. Inside a charged nanopores, besides the contact ion pair formation, the surface charges also play a significant role in reducing the counterion mobility through surface trapping. It is uncovered that the mobility of $Na^+$ ions increases first and then decreases with the surface charge density, while $Cl^-$ ions have the opposite trend. A modified first-passage time model is proposed to take into account the ion pair formation and the trapped ions inside a nanopore, which provides a clear picture in describing ion transport through a nanopore.

† Z. L. and Y. Q. contributed equally to this work

* The correspondence should be addressed to yunfeichen@seu.edu.cn

# Introduction

Porous two-dimension materials have attracted much attention in nanofluidics due to their high permeation and low resistance to ions or molecules. They have been widely used in biomolecules sensing[1], material separation[2], and energy conversion[3]. Usually, in the setup of the above applications, the porous membrane is placed between two reservoirs and nanopores on the membrane provide pathways for ions, water, and other molecules. Graphene[4] and $MoS_2$[5] nanopores have been used for DNA and protein sequencing with a super high resolution due to the atomic thickness of both materials. With the help of the surface charges on the rim of nanopores, the nanopores have strong ionic selectivity which provides a perfect platform for seawater desalination[6-7] with a low-energy cost, and for energy conversion[8] with high efficiency. Because of the dependence of underlying mechanism for all the potential application on the ionic transport, it is critical important to understand the ionic transport in nanopores.

Ionic conductance through nanopores is usually used to evaluate the ion transport quantitatively, which often deviates from bulk behavior due to various factors, including surface charge and size effects. In nature, surface charges can form at the solid-liquid interface easily due to the deprotonation/protonation of surface groups or adsorption of charged species[9]. Surface charges could attract counterions into a nanopore electrostatically, which will enhance the solution conductance compared to the bulk at low salt concentrations[10]. Besides surface charges, other surface properties have also been considered. For example, ionic rectification in conical nanopores can be tuned by adjusting the permittivity of the pore surface[11]. Fast ion transport has been founded in carbon nanotubes with sub-2 nm in diameter due to the frictionless inner surfaces[12].

Nanopores provide highly confined spaces for ions and water molecules. Therefore, surfaces have intrinsic properties in affecting the ion transport through a nanopore. With the decrease of a nanopore size, surface effects will dominate in ionic transport. For example, in a charged conical pore with a nanoscale tip and a microscale base, obvious ionic current rectification can be found[13-14]. For cylindrical pore, sub-continuum ion transport can occur, such as the nonlinear current-voltage curves[15-16] which may be difficult to distinguish whether its origination from ionic coulomb blockade[15] or dehydration effect[17]. Besides the surface effects on the ion transport, size effects will also play a critical role in affecting the ion transport. With the nanopore size decreased to several nanometers, water molecules and ions are crowded inside the nanopore. Ion-ion and ion-water interactions deviate from the case in bulk solutions. In nanochannels, Duan et al.[18] found that protons would transport faster than that in bulk cases in 2-nm-high nanochannels, which was attributed to the well-organized water structure near the channel surfaces. Geim et al.[19] found that ions would transport through angstrom-scale slits with reduced mobility, which was caused by the high energy cost in the dehydrating process when the ions entered such narrow channels. Most recently, our group showed that at high electrolyte concentrations, ion mobility in small nanopores could be significantly reduced from the corresponding bulk value which was caused by the electrostatic attraction between the surface charges and the counterions in the Stern layer, as well as enhanced pairing between partially dehydrated ions of opposite charges.[20]

In order to understand and control ion transport in nanopores, the dynamic process of ion transport in graphene nanopores was simulated extensively with molecular dynamics (MD) simulations in this study. The effects of surface charges and pore sizes on the ion mobility were investigated. With the advantage of the simulations, the duration time for each ion passing through the nanopore was analyzed. It was found that the effects of surface charges and pore sizes on the ion transport were attributed to the trapping and enhanced probability of forming contact ion pairs in the confined space. Based on this finding, a modified first-passage time model was proposed to predict the ion transport through a nanopore, in which the ion translocation was considered as one-dimensional biased diffusion process. The probability density distribution of duration time was obtained by solving the Fokker-Planck equation with proper boundary conditions[21]. The modified first-passage time model agreed well with the molecular dynamics simulations.

## Methods

Figure 1 demonstrates our MD simulation system for investigating ion transport through a graphene nanopore. The membrane with ten-layer graphene was used and the nanopore was created by removing carbon atoms that met the condition $x^2 + y^2 < (d_1/2)^2$, in which $d_1/2$ was the corresponding nominal pore radius calculated as the distance between the pore axis and the center of the carbon atoms on the pore wall. The effective pore diameter $d$ was defined as the nominal pore diameter $d_1$ minuses the van der Waals (vdW) diameter of carbon atoms. In this work, the pore diameter $d$ was considered from 1.4 nm to 6 nm. In order to maintain constant porosity (pore area/total membrane area) for each case, the inscribed circle diameter of the hexagonal prism membrane was set as from ~2.4 nm to 10 nm depending on the pore diameter. For each case, the length of the system was ~16 nm in the $z$ direction.

At the beginning, all atoms on the graphene membrane were electric neutral. Then, target surface charge densities of the nanopore were set by distributing specific amount of charges to the selected carbon atoms on the nanopore inner surface[22]. During the simulation, graphene atoms were restrained to their positions and conducted thermal vibration. The systems were solved using the Solvate plugin of VMD. Sodium and chloride ions were added to bring the ion concentration to 1 M. Extra $Na^+$ ions were also added to neutralize the system which amount depended on the charge states of the nanopores.

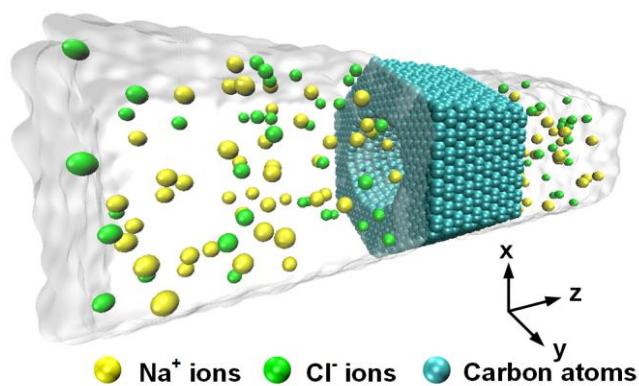

Figure 1. Scheme of the simulation system. A ten-layer graphene membrane separates two reservoirs filled with NaCl solution. A nanopore is fabricated through the graphene membrane to connect the two reservoirs. $Na^+$, $Cl^-$ ions, and carbon atoms are shown as yellow, green, and cyan spheres, respectively. Water molecules are simplified as the transparent background for clarity. A transmembrane potential along the $z$ axis is applied across the nanopore. Figure S1 in our supporting material presents the distributions of charged carbon atoms on the nanopore inner surface. The systems were built by VMD 1.9.2[23].

All simulations were conducted using NAMD 2.12[24] with a time step of 2 fs and with periodic boundary conditions in all three directions. The TIP3P water model[25] was used in this work and CHARMM36 force fields[26] were selected to describe the interactions between atoms. All carbon atoms on graphene membrane were modeled as type CA atoms[27]. A cutoff of 12 Å with the switching distance of 10 Å was set for the non-bonded interactions. Particle-Mesh-Ewald method[28] with a 1.0 Å spaced grid was used to calculate full electrostatic forces every 4 fs. The rigidity of water molecules was maintained by SETTLE algorithm[29].

For each case, the system was minimized for 0.1 ns firstly. Then the system was equilibrated for 10 ns in the NPT ensemble (constant number of particles, pressure and temperature) using the Nose-Hoover Langevin method[30] with both period and decay of 0.2 ps to keep a pressure of 1 atm. Langevin thermostat with a damping coefficient of 0.2 ps$^{-1}$ was applied to the carbon atoms and oxygen atoms to maintain a temperature of 300 K. Another 10 ns equilibration was followed to generate the initial atomic configuration in the NVT ensemble (constant number of particles, volume and temperature). An electric field $E_z$ was applied along the $z$ axis and the transmembrane voltage was calculated as[31] $V_S = -L_z E_z$, where $L_z$ was the system length in $z$ direction. The production runs varied from 100 ns to 800 ns based on the convergence of ion velocity and the number of ion permeation events for statistical analysis.

The ionic current was calculated as[31]

$$I(t) = \frac{1}{\delta t L_p} \sum_i^N q_i (z_i(t+\delta t) - z_i(t)) \qquad (1)$$

where the nanopore was located from $-L_p/2 \leq z \leq L_p/2$ along the $z$ direction, and $L_p = 3.4$ nm was the thickness of the graphene membrane. $\delta t = 2$ ps was the time interval between the consecutive frames. $z_i$ and $q_i$ were the $z$ coordinate and the charge of ion $i$, respectively.

The number of ions inside the nanopore was counted separately in each frame. Then, the total average number of ions in the nanopore was obtained by averaging over all the frames of the MD trajectory. The instantaneous velocity of each ion was calculated as $v = (z_{j+1} - z_j)/\delta t$, where $z_{j+1}$ and $z_j$ indicated the $z$ coordinates of the ions in the $(j+1)$th and $j$th frame of the trajectory. The averaged ion velocity was determined by averaging over all frames of the MD trajectory for all the ions present in the pore. The velocity of water molecules was calculated using the same method as ions. Similar to the methods described in previous studies[32-34], the electrophoretic ion mobility was determined as

$$\mu_{ion} = (v_{ion} - v_{water})/E_p \qquad (2)$$

where $v_{ion}$ and $v_{water}$ were the averaged velocity of ions and water molecules along the $z$ axis, respectively. The electrostatic potential was calculated by solving the Poisson equation using the PMEpot plugin[35] of VMD, from which one-dimensional profile of electrostatic potential along the $z$ axis was obtained. The electric-field strength inside the nanopore was calculated as $E_p = (V_{-Lp/2} - V_{Lp/2})/L_p$, where $V_{-Lp/2}$ and $V_{Lp/2}$ were the electrostatic potential at the two ends of the nanopore, respectively.

In each simulation case, the trajectory of each ion $i$ was obtained. A successful permeation event for an ion $i$ passing through the pore was determined as that the ion enters the pore from the entrance at time $t_1$ and exits the pore from the exit at time $t_2$. The translocation time of this permeation event was defined as $\Delta t = t_2 - t_1$. In the simulation, the trapping boundary was set as a parallel surface to the pore inner surface with a separation of $r_1$, where $r_1 = 0.13$ nm was the vDW radius of $Na^+$ ions. The trapping event was defined as that a successfully permeated ion stayed within the trapping boundary lasting for at least $\tau > 10$ ps during its transport process. 10 ps was determined as the time for an ion diffuse freely as long as $r_1$, i.e., a free $Na^+$ ion can diffuse ~0.15 nm with a bulk diffusion coefficient 1.33 nm$^2$/ns over a period $\tau = 10$ ps. An event of contact ion pairing (CIP) was defined as that a successful permeated ion and its counterion held together within a short distance $r_2$ instantaneously during the transport process. $r_2 = 0.25$ nm was determined as the equilibration distance between $Na^+$ ions and $Cl^-$ ions where the attractive electrostatic force equals the repulsive vdW force. $r_2$ could also be determined as the start location of $Cl^-$ ions in their radial density function around $Na^+$ ions (Fig. S2).

## Results and Discussions

In literatures, the ionic current through a nanopore should consider the access and pore resistances, as well as surface charges on the pore walls[36-39]. Figure 2 plots the comparison between the ionic currents from our MD simulations and the theoretical predictions as a function of the surface charge density. The theoretical values were calculated with the equations from (3) - (6).

$$I = Vs \cdot G = Vs \cdot \kappa_b \cdot [\frac{4L_p}{\pi d^2} + \frac{1}{d}]^{-1} \qquad (3)$$

In Eq. (3), $Vs$ is the applied voltage, $G$ is the nanopore conductance, $\kappa_b$ is the conductivity of the solution, $d$ is the diameter and $L_p$ is the length of the pore. The first and second terms represent the resistance of the cylindrical pore and the access resistance on both sides of the nanopore suggested by Hall[36]. Equation (3) has been extended by Wanunu et al.[37] in order to take into account the ion mobilities of $Na^+$ and $Cl^-$ ions inside a nanopore with the following formula:

$$\kappa_b = (\mu_{Na^+} + \mu_{Cl^-}) \cdot n_{NaCl} \cdot N_A \cdot e \qquad (4)$$

where $n_{NaCl}$ is the ion concentration of the solution, $N_A$ is the Avogadro constant, and e is the elementary charge.

Referring the equations raised by Smeets et al.[38], the surface conductance can be taken into account by the Eq. (5) as below:

$$I = Vs \cdot G = Vs \cdot [\kappa_b(\frac{4L_p}{\pi d^2} + \frac{1}{d})^{-1} + \frac{\pi d \sigma \mu_{Na^+}}{L_p}] \qquad (5)$$

where the $\sigma$ is the surface charge density on a nanopore inner surface.

Equation (6) follows Lee et al.'s model[39], which has been widely accepted because it has taken both the surface conductance and electrical access resistance into account.

$$I = Vs \cdot G = Vs \cdot \kappa_b \cdot \left[ \frac{4L_p}{\pi d^2} \frac{1}{1 + 4\frac{l_{Du}}{d}} + \frac{2}{\alpha d + \beta l_{Du}} \right]^{-1} \qquad (6)$$

Following Lee et al.'s paper, $\alpha = 2$ is a geometrical pre-factor and $\beta = 2$ is a numerical constant, and $l_{Du}$ is the Dukhin length[40].

In our simulations, $Vs = 2$ V, $d = 3$ nm, $L_p = 3.4$ nm, $n_{NaCl} = 1$ M, $\mu_{Na+} = 4.55 \times 10^{-8}$ m$^2$V$^{-1}$s$^{-1}$ and $\mu_{Cl-} = 7.53 \times 10^{-8}$ m$^2$V$^{-1}$s$^{-1}$ are the $Na^+$ and $Cl^-$ ion mobilities in 1 M bulk NaCl solutions, respectively. $\kappa_b = 11.65$ S/m is the according solution conductivity. As shown in Fig. 2, the ionic current predicted by Eq. (3) keeps unchanged due to not considering the surface charges. The ionic current predicted by Eq. (5) or Eq. (6) increases linearly with the surface charge density. As we can see, the ionic current obtained from the MD simulations deviates greatly from those equation predictions, especially at high surface charge density, which implies that it is not enough to only consider the surface charge effects and the access resistance for nanopores with diameter below 3 nm. With the increase of the surface charge densities, Fig. 2 also demonstrates that the Na$^+$ ion current increases first and then decreases at a turning point ~-0.1 C/m$^2$, and the Cl$^-$ ion current has the opposite trend. This phenomenon cannot be captured by equations (3, 5, 6). In order to confirm this is a general phenomenon for small nanopores, we have also simulated nanopores with diameter of 2 nm and 4 nm. Both the nanopores with diameter of 2 nm and 4 nm demonstrate the similar trends as displayed in Fig. S3 in the supporting material.

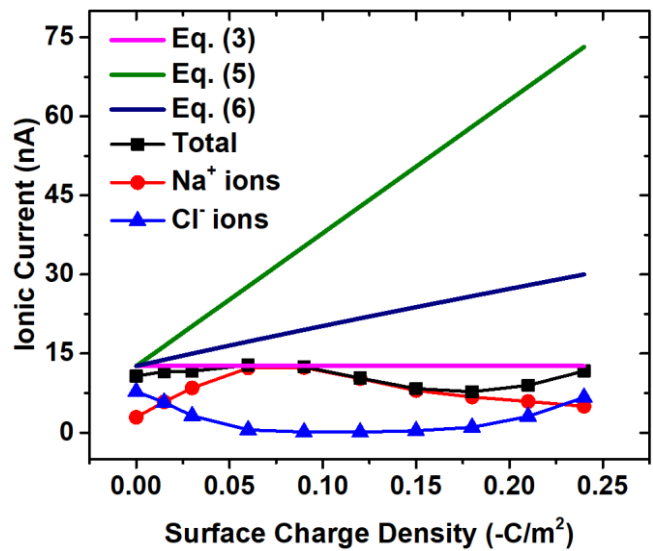

Figure 2. Comparison between the MD simulation results and the theoretical predictions from equations (3, 5, 6). Ionic currents from the MD simulations and the theoretical predictions were obtained in a 3-nm-diameter nanopore under different surface charge densities ranging from -0.24 to 0 C/m$^2$.

In order to understand the deviation between the simulation results and theoretical predictions, we analyzed the number and mobility of ions inside the nanopore. As shown in Fig. 3(a), the number of Na$^+$ ions inside the nanopore increases with the surface charge density due to the electrostatic attraction from the negative charges on the nanopore inner surface. However, the number of Cl$^-$ ions decreases first with the surface charge density due to the electrostatic repulsion, and increases after the transition point at ~-0.1 C/m$^2$ because charge inversion[41] occurs and Cl$^-$ ions start to act as counterions in order to screen the inverted surface charges. At solid-liquid interface, the correlation among ions cannot be neglected especially in the cases with high surface charge densities and high salt concentrations[42]. As plotted in Fig. S4, $Na^+$ ions are overcrowded near the highly charged surfaces. Once the charge inversion occurs, $Cl^-$ ions act as counterions to screen the inverted surface charges resulting in a higher concentration than $Na^+$ ions in the center of the pore. From the ion distributions (Fig. S5), $Na^+$ ions mainly aggregate in the vicinity of the charged surface, while $Cl^-$ ions barely exist in this region and mainly locate in the center region of the nanopore. These phenomena have been addressed in our previous work[43]. Here we note that in Fig. 3(a) the total number of ions inside the nanopore keeps almost constant with the surface charge density increasing from 0 to -0.1 $C/m^2$. For weakly charged nanopores, the similar ionic current from simulations could be explained by the almost unchanged total ion number in the nanopore. However, with the surface charge densities increasing above -0.12 $C/m^2$, the total ion number inside the nanopore increases obviously with the surface charge density, which shares the same trend as that of the ionic current predicted by Eq. (5) and Eq. (6) while has a different trend from that of the MD simulation. Thus the large deviation between the MD simulations and the continuum theory cannot be explained from the number of ions inside the nanopores. Therefore, we focused on the ion mobility when ions transport in the nanopore.

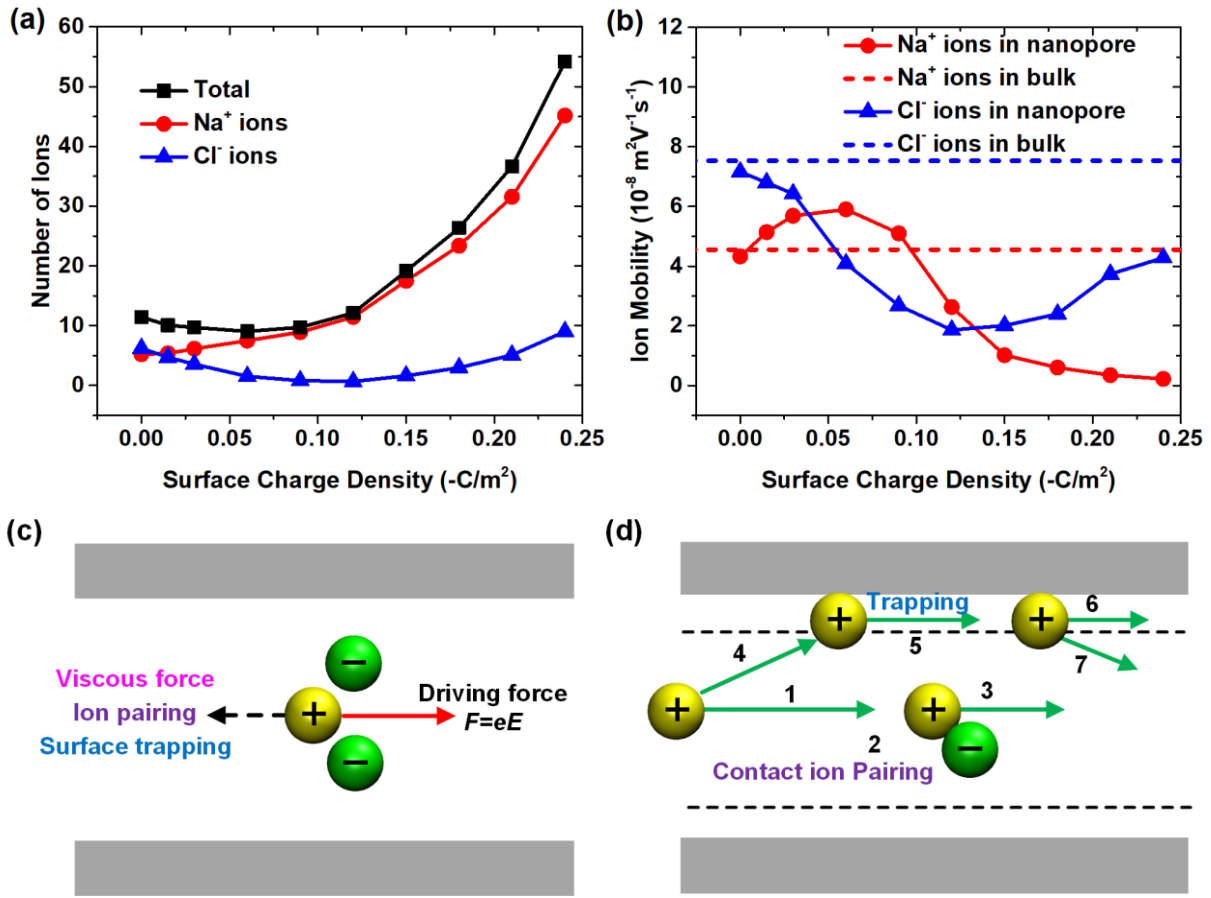

Figure 3. Ion number and mobility (a and b), and schematic of the ion transport process (c and d) in nanopores. (a) The average number of Na$^+$ ions, Cl$^-$ ions and total ions in a 3-nm-diameter nanopore as a function of the surface charge density; (b) Ion mobility of Na$^+$ ions and Cl$^-$ ions in the 3-nm-diameter nanopore as a function of surface charge density; (c) Schematic of the forces that an ion experiences when moving in the nanopore; (d) Schematic of the trajectories and events that may occur when an ion transports through the nanopore. The gray rectangle represents the surface boundary of the nanopore and the dash line represents the trapping boundary. The corresponding bulk value of ion mobility was calculated in 1 M NaCl bulk electrolyte under an electric field of 0.3 V/nm, which approximates the electric-field strength inside the nanopore.

Figure 3(b) plots the ion mobility of Na$^+$ and Cl$^-$ ions in 3-nm-diameter nanopores as a function of the surface charge density. It is found that with the increase of the surface charge density, the mobility of Na$^+$ ions increases first over the bulk value and then decreases severely at high surface charge densities. The mobility of Cl$^-$ ions is always smaller than their bulk value, decreasing with the increase of the surface charge density until at the transition point of ~-0.1 C/m² and then starting to increase. Similar phenomena are also observed in another two nanopores with 2 nm and 4 nm in diameter (Fig. S6). The classical Debye and Hückel theory[44] indicated that ion mobility decreases as concentration increases. As displayed in Fig. 3(a), the total ion number keeps an increase trend, from which it is estimated that both the Na$^+$ and Cl$^-$ ion mobility should decrease with the increase of the surface charge density. However, this is different from our MD results, in which Na$^+$ ion mobility increases first then decreases while the Cl$^-$ ion has the opposite trend with the increase of the surface charge density. Here we studied the forces that ions might experience when transporting in small nanopores to uncover the factors that influence the ion mobility. As shown in Fig. 3(c), ions are driven by an external electric field along the *z* direction and are dragged back by the viscous force from the water molecules, the electrostatic forces from counterions and the surface charges. The electric-field strength lies within the linear response regime. As depicted in Fig. 3(d), when a Na$^+$ ion enters the pore driven by the external electric field, it may encounter a coming counterion or be trapped by surface charges. Of course, some ions may directly transport through the nanopore without interacting with any other ions or surface charges. In this case, the ion mobility is only affected by the viscous force from the water molecules. This part ions should obey the classical Debye and Hückel theory that their mobility is only dependent on the salt concentration. Now we focus on the events of ion trajectory 1-2-3 and 4-5-6, which have long duration time in translocating through the nanopore. When the diameter of a nanopore is reduced to below 3 nm, part of hydration layer should be peeled off from the ion when it enters the nanopore. Once the hydration layer around ions is broken, it is easier to form ion pairs than those ions screened by complete hydration layers, which reduces the ion mobility greatly. Another mechanism in reducing the ion mobility is attributed to the electrostatic trapping by the surface charges as shown in Fig. 3(d).

It should be noticed here, besides the ion pair formation and the surface trapping, the high viscous water may also slow down the ion mobility inside a nanopore. With the decrease of the nanopore diameter, the effective water viscosity may increase[40]. However, through careful analysis, it is found that the effect of the increased water viscosity on reducing ion mobility can be neglected compared with that of the ion pair formation or of the surface trapping ion. As reported in literatures[45], the effective viscosity of the first water layer near the wall is about 20% higher than that in the bulk, and the viscosity in the other region away from the wall keeps unchanged. For an uncharged nanopore, ions mainly distribute in the center of the nanopore. In this case, the increased water viscosity in the Stern layer affects little on the transport of the ions because ions are mainly distributed in the center of the nanopore. If the nanopore has charged inner surfaces, the surface charges may attract a large amount of counterions to form the electrical double layers. In this case, surface trapping plays the dominant role in reducing the ion mobility. So, we only consider the effect of the ion pairing and surface trapping on ion transport inside a nanopore. The ion transport process can be categorized as direct translocation events, CIP events, and trapping events using our definition in the analysis protocol. The contact ion pair is one type of the ion pair[46], which means that ions are in direct contact without water medium between them. According to the conventional Stern electric double layer model, the trapping boundary used here corresponds to the inner Helmholtz layer[47].

Having this physical picture of the forces acting on each ion in mind, the factors affecting ion mobility through a nanopore are mainly attributed to the nanopore size and the surface charge density. With the advantage of MD simulations, we can create neutral nanopores artificially. For the size effect, Fig. 4 shows the ion transport behaviors in uncharged nanopores with different diameters. As shown in Fig. 4(a), the ion mobility of both $Na^+$ and $Cl^-$ ions in the pore is smaller than the corresponding bulk value when the pore diameter is below 5 nm, and the ion mobility will get smaller with the decrease of the pore size. The number ratio of each type of ions over the total number of ions keeps unchanged with the nanopore size (Fig. S8)[48]. In order to find the reason for the reduced ion mobility in small nanopores, dwell time distribution of ions is analyzed as shown in Fig. 4(b), imitating the data analysis of a DNA strand translocating through a nanopore[49-57]. The dwell time distribution characterizes the time it costs when a $Na^+$ ion transport through the nanopore. It is observed that the dwell time distribution of $Na^+$ ions in a 1.4-nm-diameter nanopore has a long tail, which indicates that some ions translocate through the nanopore slowly. As the pore size gets smaller, longer translocation events are observed. By analyzing the contact ion pairs between $Na^+$ and $Cl^-$ ions, the long dwell time is found due to the appearance of the CIP events, which contribute significantly to the mobility reduction of $Na^+$ ions.

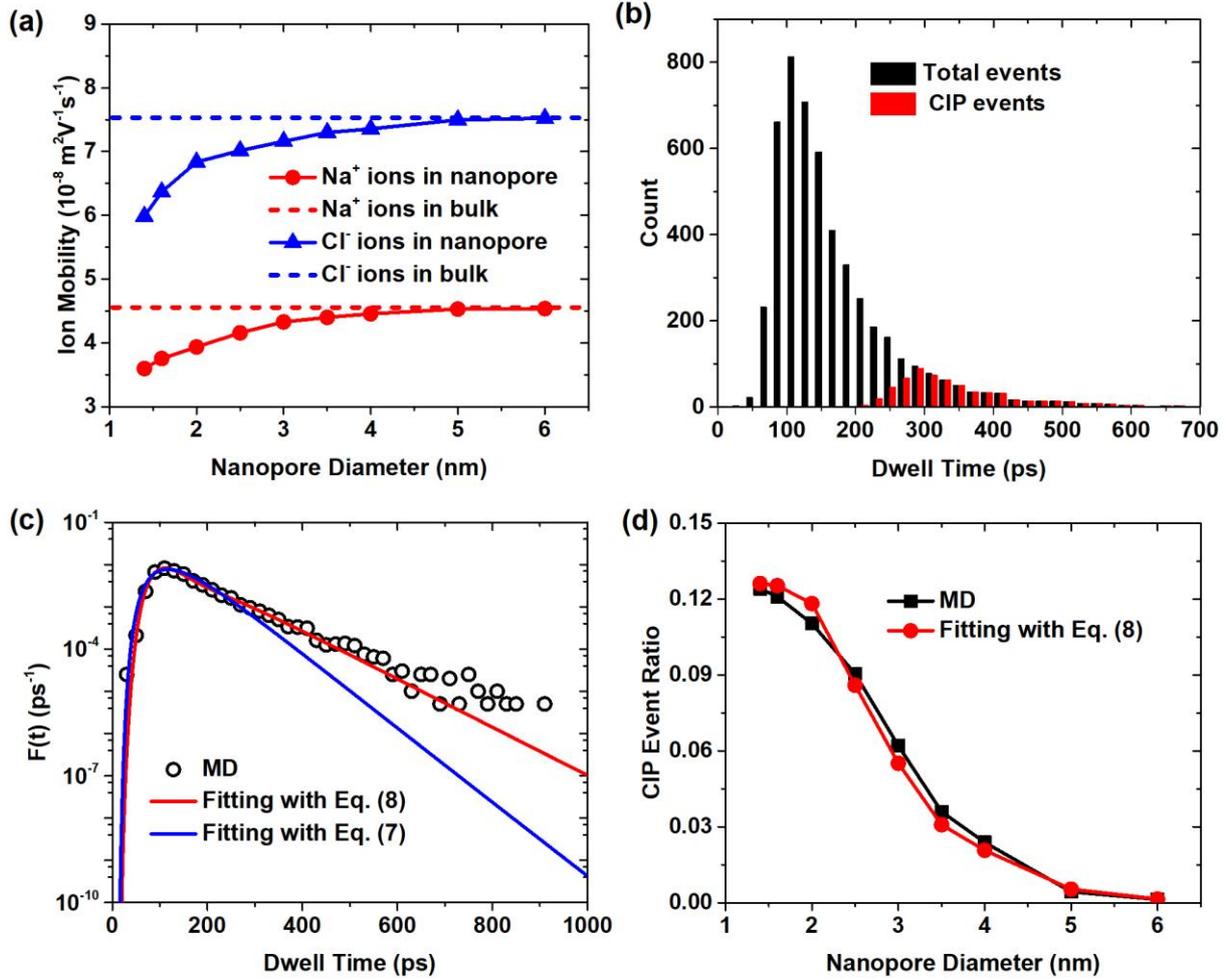

Figure 4. Ion transport behavior in uncharged nanopores with different diameters. (a) Ion mobility of $Na^+$ and $Cl^-$ ions in uncharged nanopores with different diameters; (b) Histogram distribution of the total events and CIP events of $Na^+$ ions when translocating through an 1.4-nm-diameter nanopore; (c) The dwell time probability distribution of $Na^+$ ions translocating through an 1.4-nm-diameter nanopore in a logarithmic view, the red and blue solid lines are the theoretical fitting with Eq. (7) and Eq. (8), respectively; (d) The ratio of the CIP events to the total ion translocation events through the nanopore from the MD simulations and the extracted ratio when fitting the dwell time probability distribution using Eq. (8) versus nanopore diameters.

The first-passage model was used to fit the dwell time distribution of ion transport events. One typical solution to the first-passage model is obtained as Eq. (7) by using the semi-infinite boundary condition[21, 53, 55, 58].

$$F(t) = \frac{L_p}{\sqrt{4\pi Dt^3}} e^{-\frac{(L_p - tv_{ion})^2}{4Dt}} \tag{7}$$

where $F(t)$ is the first-passage probability density function, which has the physical meaning of the probability per unit time for the ion passing through the pore. $D$ and $v_{ion}$ are the diffusion constant and the drift velocity of the ions inside the pore, respectively. As shown in Fig. 4(c), the overall agreement between Eq. (7) and the MD results is good for the short dwell time. From the above analysis of the CIP events, the disagreement between the prediction and the long dwell time can be explained by the existence of the CIP events when cations and anions transport in small pores, resulting in two different types of translocation events. Denoting the probabilities of direct translocation events and CIP events by $p$ and $(1-p)$ and the corresponding two distributions by $F_1(t)$ and $F_2(t)$, the total distribution $F(t)$ can be described by an extended model Eq. (8).

$$F(t) = pF_1(t) + (1-p)F_2(t) \tag{8}$$

Within this extended model, $F_1(t)$ and $F_2(t)$ are both in the form of Eq. (7). And for both of them, the value $D/v_{ion}$ should be the same according to the Einstein relationship[59-61]. As shown in Fig. 4(c), the extended theory agrees well with the MD results. The ratio of the CIP events to the total ion translocation events were also analyzed with different pore sizes as shown in Fig. 4(d). With the decrease of nanopore size, more percentage of the CIP events happens. The CIP event ratio is close to zero when the pore size is larger than 5 nm and reached to 12% when the pore is 1.4 nm in diameter. The increased CIP event ratio can be used to explain the reduced ion mobility for both the cations and anions inside the nanopore. The ratio values obtained from the MD simulations and the theoretical model are in good agreement, which indicates that two groups of ion translocation events could be identified using Eq. (8). Position dependent ion mobility[62] can also confirm the effect of CIP. As shown in Fig. S9, the ion mobility of both $Na^+$ and $Cl^-$ ions decreases with the nanopore shrinking, especially in the pore center region, which is caused by the more CIP events when the confined effect becomes strong. We have also analyzed the CIP event ratio with salt concentration (Fig. S10). With the increase of the salt concentration, the ratio of the CIP events increases, which reduces the ion mobility.

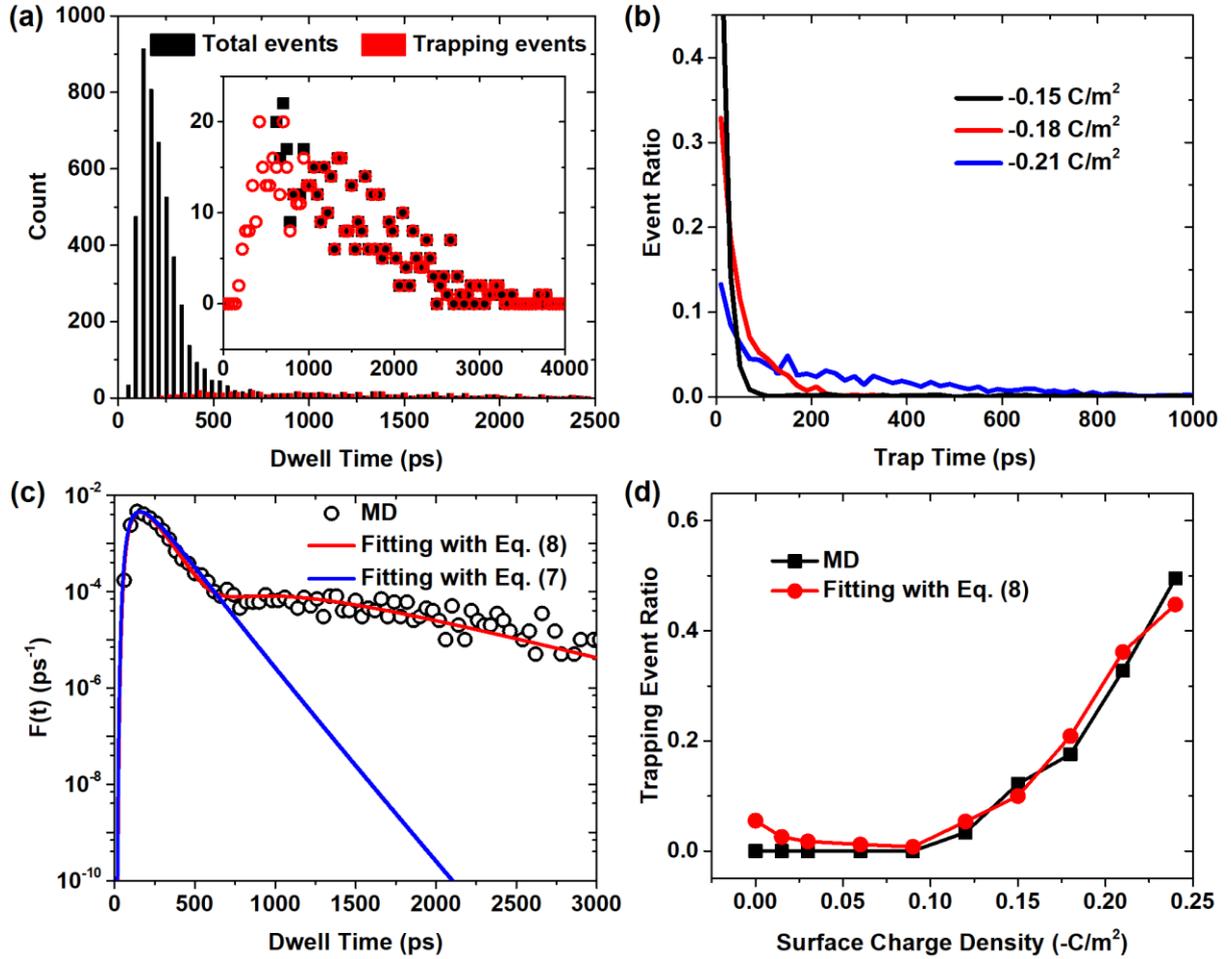

Figure 5. Ion transport behaviors in a 3-nm-diameter nanopore with different surface charge densities. (a) Histogram translocation distribution of total events and the trapping events of Na$^+$ ions at a surface charge density of -0.15 C/m$^2$, and the inset shows an enlarged view of the distribution tails of the two type events; (b) The trapping time ratio distribution of the trapping events at the surface charge density of -0.15 C/m$^2$, -0.18 C/m$^2$ and -0.21 C/m$^2$, respectively. (c) The dwell time probability distribution at a surface charge density of -0.15 C/m$^2$ in a logarithmic view, the solid lines are the theoretical fitting according to Eq. (7) and Eq. (8), respectively; (d) The trapping events ratio from the MD simulations and the extracted ratio when fitting the dwell time probability distribution using Eq. (8) versus surface charge density.

To take the surface charges into consideration, we also investigated the ion transport behaviors in 3-nm-diameter nanopores with different surface charge densities as shown in Fig. 5. The dwell time distribution of Na$^+$ ions translocating through the 3-nm-diameter nanopore with a surface charge density of -0.15 C/m$^2$ was shown in Fig. 5(a) with a visible tail. Under such a high surface charge density, the inner Helmholtz layer[63] appears (Fig. S5) caused by the strong electrostatic interactions between ions and surface charges, which indicates that the ions are trapped during their translocation process. To quantify this behavior, we analyzed the translocation events in ion trajectories, and found that the trapping events were responsible to the long tail in the dwell time distribution.

The trapping time $\tau$ was counted when the ions had a distance below 0.13 nm from the surface. Figure 5(b) plots the ratio distribution of the trapping time at the surface charge density of -0.15 C/m$^2$, -0.18 C/m$^2$ and -0.21 C/m$^2$, respectively. We can find under a higher surface charge density, an ion will experience a longer trapping time with larger probability. This also indicates that when ions are trapped, it is hard for them to be released at high surface charge density. As shown in Fig. 5(c), the dwell time distribution was fitted with Eq. (7) and (8). Equation (7) predicted the MD results well except the long dwell time. The disagreement in this region with long dwell time is attributed to the occurrence of trapping events when ions transport in highly charged nanopores. With the consideration of the surface trapping, the extended theory predicted the simulation data well, which reveals that the translocation events might be categorized as two groups with well-separated time scales.

The trapping ratio in the simulations was analyzed in Fig. 5(d). The value obtained from the MD simulation and fitting results have a good agreement, which indicates that the two type ion translocation events could be identified using Eq. (8). We note that a little disagreement at zero surface charges is due to the presence of the CIP events at this condition. When the surface charge density was above -0.1 C/m$^2$, trapping events happened much more frequently. The trapping events ratio was ~10% at a surface charge density of -0.15 C/m$^2$ and can reach to ~50% when the pore was charged to -0.24 C/m$^2$. The increased trapping events ratio at high surface charge density could be the indicative of strongly reduced mobility.

After the CIP events and trapping events were found during the translocation of ions through charged nanopores, the behavior of ion mobility shown in Fig. 3(b) can be explained. Position dependent ion mobility[62] can also confirm the effects of surface trapping and CIP (Fig. S11). With the surface charge increasing from 0 to -0.06 C/m$^2$, Na$^+$ ions mobility increases. This is attributed to less interactions between Na$^+$ ions and Cl$^-$ ions. At low surface charge density, there are no trapping events for Na$^+$ ions (Fig. 5), and the ratio of Na$^+$ ions increases at low surface charge density (Fig. S7) which means that Cl$^-$ ions pose weak interactions on the moving Na$^+$ ions due to the decreasing number of Cl$^-$ ions with the surface charge density. The ion mobility of Na$^+$ ions increases to even over the bulk mobility. For Cl$^-$ ions, they mainly distribute in the center of the pore. Na$^+$ ions pose large drag resistance on the Cl$^-$ ions with the increase of the Na$^+$ ions, which results in the decreased mobility of the Cl$^-$ ions. As the surface charge density increases further, surface trapping caused by the attraction between the surface charges and Na$^+$ ions becomes more obvious and the ion mobility of Na$^+$ ions decreases. Under strongly charged cases, where charge inversion[41] happens near the surface (Fig. S4), Cl$^-$ ions act as the effective counterions and their mobility increases, which is similar to the trend of Na$^+$ ions mobility under low surface charge density.

## Conclusions

Through investigation of ion transport with MD simulations, ion mobility can be regulated via the nanopore sizes and surface charges. In neutral or weakly charged nanopores, the CIP events decrease the ion mobility, which becomes more obvious in highly confined spaces. As the surface charge density increases, the electrostatic interactions between the surface charges and counterions get stronger. Then, surface trapping effect on counterions will decrease the counterion mobility. In order to fit the dwell time of ions from the simulation, an extended first-passage equation was raised which completely covers the broad dwell time distribution of ion translocation time. The proposed model can well predict the diverse ion translocation events through a nanopore, which can probably be applied to predict the translocation of other molecules through a nanopore. The extended model provides important insights into ion or molecule transport in a nanofluidic device.

## Acknowledgement

The authors thank the Natural Science Foundation of China (Grant No. 51435003). Zhongwu Li is supported by the Scientific Research Foundation of Graduate School of Southeast University (Grant No. YBJJ1802) and the Postgraduate Research & Practice Innovation Program of Jiangsu Province (Grant No. KYCX18_0067), and acknowledges financial support from the China Scholarship Council (CSC 201806090020).

## Supporting Information

Figure S1: Charge distributions on the nanopore inner surface for nanopores with different diameters. Figure S2: The radial density distribution of $Cl^-$ ions around $Na^+$ ions in bulk solutions. Figure S3: Comparison between simulation results and theoretical prediction in nanopores with diameters of 2 nm and 4 nm. Figure S4: Charge inversion phenomenon in highly charged nanopores. Figure S5: Ion distributions in 3-nm-diameter nanopores. Figure S6: Ion mobility in nanopores with diameter of 2 nm and 4 nm. Figure S7: The ratio of different ions in 3-nm-diameter nanopores. Figure S8: The ratio of different ions in uncharged nanopores with different diameters. Figure S9: Ion mobility in uncharged nanopores of different diameters along the radial direction. Figure S10: CIP event ratio of $Na^+$ ions during transporting in bulk and in a nanopore with different salt concentrations. Figure S11: Ion mobility in 3-nm-diameter nanopores along the radial direction under different surface charge densities.

## Author Information

**Corresponding Author**

*E-mail: yunfeichen@seu.edu.cn. Tel: 86-13815888816 (Y.C.).

**Author Contributions**

Z.L. and Y.Q. contributed equally to this work.

**Notes**

The authors declare no competing financial interest.

**Figure for Table of Contents**

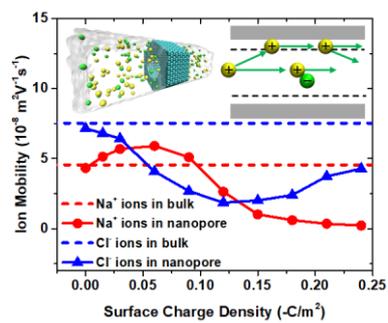

Table of Contents graphic showing the ion mobility as a function of surface charge density along with the schematic of surface trapping and contact ion pairing in highly confined nanopores.